\documentclass[aps,pre,preprint,floatfix,twoside,tightenlines,showpacs,
showkeys]{revtex4}
\usepackage{amsmath,graphicx}
\begin{document}
%
%--------------------------------------
\title{Generalized Lattice Model of Multi-Component Systems with Internal Degrees of Freedom. II. Quasiequilibrium States}
\author{A.Yu.~Zakharov}\email[E-mail: ]{Anatoly.Zakharov@novsu.ru}
\author{M.I.~Bichurin}\email[E-mail: ]{Mirza.Bichurin@novsu.ru}
\affiliation{Novgorod State University, Veliky Novgorod, 173003, Russia}
%
%
%----------------------------------------------------------------------
\begin{abstract}
The paper contains an application of the generalized lattice model to multicomponent systems with internal degrees of freedom. The short-range inter-atomic repulsions and smooth long-range  parts of the inter-atomic potentials are considered separately by means of packing condition and in effective field approximation, respectively.  The dependence of the inter-atomic potentials on the internal degrees of freedoms (such as atomic electric and/or magnetic momentum) taken into account. The Helmholtz free energy functional in the generalized lattice model is reduced to the Ginzburg-Landau-Cahn-Hilliard-like (GLCH) form. The connection between the inter-atomic potentials characteristics and the parameters of the GLCH-like functional is obtained. Equations for both equilibrium and quasi-equilibrium states in condensed systems are derived. It is shown that equilibrium distribution of the fast internal degrees of freedom by frozen space distribution of the components obeys to the Schr\"odinger-like equation.
\end{abstract}
%------------------------------------------------------------------------------
%
\maketitle
%------------------------------------------------------------------------------

%------------------------------------------------------------------------------
\section{Introduction}

The generalized lattice model (GLM) of multicomponent condensed systems (such as solid or liquid solutions) was proposed in paper~\cite{ZT} and developed in~\cite{ZZL,ZL,Z1,Z2}. In contradistinction to usual lattice models (see for example~\cite{Prig,Smir,Gray,Khach,Israel}), the GLM takes into account the following essential factors: 

\begin{enumerate}
\item  The short-range inter-atomic repulsions. These repulsions are not identical for different pair of atoms, therefore it is impossible to take into account the repulsions by means of lattice introduction.

\item  The presence of the local fields due to the long-range parts of inter-atomic potentials. These fields have the essential influence on both equilibrium properties and non-equilibrium processes on the corresponding space scales.

\item  The Helmholtz free energy functional contains the well defined parameters that have connections with characteristics of the components and their interactions. 
\end{enumerate}

The present paper contains further development of GLM. In addition to previous results it takes into account the following: 

\begin{enumerate}
\item  Existence of the internal atomic degrees of freedom such as atomic electric and magnetic moments. These degrees of freedom are responsible for the local magnetization and local electric polarization in the system. 

\item  Existence of some non-equilibrium degrees of freedom due to colossal time of their relaxations. Real condensed systems are, as a rule, partially equilibrium systems.

\item  Existence of essential inhomogeneities due to both difference of the atomic sizes of the components and presence of some frozen degrees of freedoms. It is well known, the inhomogeneities scales can be arbitrary up to nanocluster sizes.
\end{enumerate}

\section{Separation of Short-Range and Long-Range Parts in the Inter-atomic Potentials}

Short-range and long-range parts of the inter-atomic potentials in condensed matter play essentially different roles. Short-range repulsions prevent the collapse of particles into multi-particle conglomerates and lead to some restriction on the local densities $n_{i} (r)$ of the components 
\begin{equation}\label{dens}
    n_i(\mathbf{r}) \le  \frac{1}{\omega_i},
\end{equation}
where $\omega _{i} $ is the inverse value of the maximal local density of $i$-th component ($i=1\div m$, $m$ is the number of the components in the system). The quantity $\omega _{i} $ has dimensionality of volume and henceforth will called as the specific atomic volume of $i$-th component. As far as the quantity $\omega _{i} {\kern 1pt} n_{i} (r)$ is the local volume fraction of $i$-th component at the point $r$, then we have the restrictions on the local densities of the components for all points $r$ in the system: 
\begin{equation}\label{pack-0}
    \sum_{i=1}^m\ \omega_i\ n_i(\mathbf{r}) \leq 1.
\end{equation}
For condensed matter the inequality in this condition should be replaced on the equality: 
\begin{equation}\label{packing}
    \sum_{i=1}^m\ \omega_i\ n_i(\mathbf{r})-1=0.
\end{equation}
This constrain (the packing condition) of the components local densities in condensed matter takes into account the short-range inter-atomic repulsions. It will be used in generalized lattice model (GLM) instead of the hypothesis on the lattice structure existence. The Helmholtz free energy minimization for equilibrium state of systems should be realized under the packing condition, which takes into account the short-range parts of the inter-atomic potentials in the system. Hence the inter-atomic potentials should be included into the free energy with cutting out their short-range parts:
\begin{equation}\label{long-range}
    K_{ij}(\mathbf{r})=\left\{%
\begin{array}{ll}
    W_{ij}(\mathbf{r}), & \hbox{if} \  |\mathbf{r}|\geq a_{ij}, \\
    0, & \hbox{otherwise,} \\
\end{array}%
\right.
\end{equation}
where  $W_{ij}(\mathbf{r})$ is ``true'' interaction potential between $i$-th and  $j$-th components, $a_{ij}$ are the cutting parameters, related to the specific atomic volumes of the components by the relations
\begin{equation}\label{cut}
    a_{ij}\simeq \left[ (\omega_i)^{1/3}+
(\omega_j)^{1/3}\right].
\end{equation}
In addition of the packing condition~(\ref{packing}), the numbers of the components atoms in the system should be fixed at the free energy minimization. These conditions have the following form:
\begin{equation}\label{N}
    \int\limits_{(V)}\ n_i(\mathbf{r})\ d\mathbf{r}-N_i=0, \qquad i=0\div m.
\end{equation}
The packing condition~(\ref{packing}) and condition of numbers particles conservation~(\ref{N}) should be satisfied for any form of the configuration part of the free energy.

\section{Helmholtz Free Energy in the Generalized Lattice Model}
The Helmholtz free energy functional of a system with account of the atomic electric~$\mathbf{D}_i \left( \mathbf{r} \right)$ and magnetic moments~$\mathbf{M}_i \left( \mathbf{r} \right)$  of the components in presence of the external electric~$\mathbf{E} \left( \mathbf{r} \right)$  and magnetic~$\mathbf{H} \left( \mathbf{r} \right)$  fields has the following form:
\begin{equation}\label{helmholtz}
\begin{array}{r}
 {\displaystyle   F = \frac{1}{2}\sum_{i,j=1}^{m}\, \iint\limits_{(V)}
    K_{ij}(\mathbf{r}-\mathbf{r}') n_i(\mathbf{r})
    n_j(\mathbf{r}') d\mathbf{r}\, d\mathbf{r}' } \\
{\displaystyle + \frac{1}{2}\sum_{i,j=1}^{m}\, \iint\limits_{(V)}
    Q_{ij}(\mathbf{r}-\mathbf{r}') n_i(\mathbf{r})
    n_j(\mathbf{r}') \left( \mathbf{D}_i(\mathbf{r})\cdot
    \mathbf{D}_j(\mathbf{r}') \right) d\mathbf{r}\, d\mathbf{r}' } \\
{\displaystyle + \frac{1}{2}\sum_{i,j=1}^{m}\, \iint\limits_{(V)}
    R_{ij}(\mathbf{r}-\mathbf{r}') n_i(\mathbf{r})
    n_j(\mathbf{r}') \left( \mathbf{M}_i(\mathbf{r})\cdot
    \mathbf{M}_j(\mathbf{r}') \right) d\mathbf{r}\, d\mathbf{r}' }\\
{\displaystyle + \frac{1}{2}\sum_{i,j=1}^{m}\, \iint\limits_{(V)}
    S_{ij}(\mathbf{r}-\mathbf{r}') n_i(\mathbf{r})
    n_j(\mathbf{r}') \left( \mathbf{D}_i(\mathbf{r}) \cdot
    \mathbf{M}_j(\mathbf{r}') \right) d\mathbf{r} d\mathbf{r}'  }\\
{\displaystyle + \sum_{i=1}^m\, \int\limits_{(V)} \left( \mathbf{E}(\mathbf{r}) \cdot
\mathbf{D}_i(\mathbf{r}) \right) n_i(\mathbf{r})\,  d\mathbf{r} +
\sum_{i=1}^m\, \int\limits_{(V)} \left( \mathbf{H}(\mathbf{r}) \cdot
\mathbf{M}_i(\mathbf{r}) \right) n_i(\mathbf{r})\,  d\mathbf{r} +}\\
{\displaystyle   + T\sum_{i=1}^m\, \int\limits_{(V)} n_i(\mathbf{r}) \ln
    \left(\frac{n_i(\mathbf{r})}{n(\mathbf{r})}\right)
    d\mathbf{r}},
\end{array}
\end{equation}
where $K_{ij} \left( \mathbf{r} \right) $, $Q_{ij} \left( \mathbf{r} \right) $, $R_{ij} \left( \mathbf{r} \right)$ and $S_{ij} \left( \mathbf{r} \right)$  are the long-range parts of the relevant two-body inter-atomic potentials related to moments independent, electric dipole -- electric dipole, magnetic dipole -- magnetic dipole and magnetoelectric terms, respectively, the last term in this relation is the entropy term, $T$ is the temperature in energetic units, $n \left( \mathbf{r} \right)$  is the summarized local density of the particles
\begin{equation}\label{density}
    n \left( \mathbf{r} \right) = \sum_{j=1}^m n_j \left( \mathbf{r} \right)
\end{equation}

The equilibrium distributions of the components in the space are determined by minimum of the Helmholtz free energy functional with account of the conditions~(\ref{packing}) and (\ref{N}). But beyond these conditions, some connections between electric $\mathbf{D}_i(\mathbf{r})$ and magnetic $\mathbf{M}_i(\mathbf{r})$  moments and the external fields should be included. Suppose that external fields influence on the orientations of related moments but do not influence on their magnitudes $D_i$  and $M_i$. Hence we have 
\begin{equation}\label{D-M}
    \left(\mathbf{D}_i\left(\mathbf{r}  \right)\right)^2 - D_i^2 = 0; \quad     \left(\mathbf{M}_i\left(\mathbf{r}  \right)\right)^2 - M_i^2 = 0.
\end{equation}
\section{The Lagrange functional}
For the minimization of the Helmholtz free energy~(\ref{helmholtz}) at the conditions~(\ref{packing}), (\ref{N}), (\ref{D-M}), let us introduce the Lagrange functional $\mathcal{L}$ depending on the local densities $n_i(\mathbf{r})$ of the components, their electric and magnetic moments $\mathbf{D}_i(\mathbf{r})$, $\mathbf{M}_i(\mathbf{r})$, the external fields $\mathbf{E}(\mathbf{r})$, $\mathbf{H}(\mathbf{r})$, and the Lagrange multipliers $\Psi(\mathbf{r})$, $\lambda_i(\mathbf{r})$, $\nu_i(\mathbf{r})$, $\mu_i$:
\begin{equation}\label{lagrange}
\begin{array}{r}
{\displaystyle
    \mathcal{L} = F - \sum_{i=1}^{m}\mu_i\left[ \int\limits_{(V)}\
    n_i(\mathbf{r})\ d\mathbf{r}-N_i \right]
- \sum_{i=1}^m\ \int\limits_{(V)} \frac{\lambda_i\left( \mathbf{r} \right)}{2} \, n_i^2\left(\mathbf{r} \right)\, \left[ \left(\mathbf{D}_i\left(\mathbf{r}  \right)\right)^2 - D_i^2 \right]\, d\mathbf{r}\ 
}\\ 
{\displaystyle -
\sum_{i=1}^m\ \int\limits_{(V)} \frac{\nu_i\left( \mathbf{r} \right)}{2} \, n_i^2\left(\mathbf{r} \right)\, \left[ \left(\mathbf{M}_i\left(\mathbf{r}  \right)\right)^2 - M_i^2 \right]\, d\mathbf{r}  - \int\limits_{(V)}\Psi(\mathbf{r}) \left(
\sum_{i=0}^m\ \omega_i\
    n_i(\mathbf{r})-1 \right) d\mathbf{r},}
\end{array}
\end{equation}
where $F$ is the Helmholtz free energy defined by~(\ref{helmholtz}). 

The necessary condition of the Helmholtz free energy extremum is vanishing of the partial and functional derivatives of the Lagrange functional with respect to $n_i(\mathbf{r})$, $\mathbf{D}_i(\mathbf{r})$, $\mathbf{M}_i(\mathbf{r})$, $\Psi(\mathbf{r})$, $\lambda_i(\mathbf{r})$, $\nu_i(\mathbf{r})$, $\mu_i$. Calculations of these derivatives lead to a system of nonlinear integral equations with kernels $K_{ij}\left(\mathbf{r} - \mathbf{r}'  \right)$, $Q_{ij}\left(\mathbf{r} - \mathbf{r}'  \right)$, $R_{ij}\left(\mathbf{r} - \mathbf{r}'  \right)$, $S_{ij}\left(\mathbf{r} - \mathbf{r}'  \right)$. Unfortunately, at present there are no effective methods of such kind equations solutions with kernels of general form. But under the some conditions this system of equations can be reduced to a system of partial differential equations. Instead of the interatomic potentials $K_{ij}\left(\mathbf{r}   \right)$, $Q_{ij}\left(\mathbf{r}  \right)$, $R_{ij}\left(\mathbf{r} \right)$, $S_{ij}\left(\mathbf{r} \right)$ this system of partial differential equations contains a set of integral characteristics of the potentials.

\section{Reduction of the Generalized lattice model to Ginzburg-Landau-Cahn-Hilliard-like approximation}
One of the most effective methods in statistical thermodynamics of condensed matter based on the phenomenological Ginzburg-Landau-Cahn-Hilliard (GLCH) models~\cite{LL,Cahn}. Unfortunately, prognostic capabilities of the GLCH models are restricted by absence of the direct connections between model parameters and inter-atomic potentials.

The GLM contains the inter-atomic potentials in the explicit form, but the mathematical structure of the free energy functional for GLM is more complicate than in GLCH model. By some additional assumptions, the GLM can be reduced to the GLCH-like model.

There are at least three scales of the sizes in the system:
\begin{enumerate}
    \item atomic sizes $a_0$;
    \item range of actions of long-range parts of the inter-atomic potentials $r_0$;
    \item distances $b_0$ on which changes local compositions and/or local moments of the components in the system.
\end{enumerate}
Suppose these parameters obey the inequalities:
\begin{equation}\label{arb}
    a_0\lesssim r_0 \ll b_0.
\end{equation}

Let us expand the functions $n_j \left( \mathbf{r}' \right) $, $\mathbf{D}_j \left( \mathbf{r}' \right) $, $\mathbf{M}_j \left( \mathbf{r}' \right) $  in vicinity of point $\mathbf{r}$   in powers of $\mathbf{r} - \mathbf{r}' $  up to second order terms. Using this expansion and the Green formula 
\begin{equation}\label{Green}
    \int\limits_{(V)}\, u\left( \mathbf{r} \right) \Delta v\left( \mathbf{r} \right)\,d\mathbf{r} = - \int\limits_{(V)}\, \left(  \nabla u\left( \mathbf{r} \right)\cdot \nabla v\left( \mathbf{r} \right) \right) d\mathbf{r},
\end{equation}
that is valid if functions $u\left( \mathbf{r} \right)$, $v\left( \mathbf{r} \right) $ and their gradients $\nabla u\left( \mathbf{r} \right)$, $\nabla v\left( \mathbf{r} \right)$  vanish on the boundary of the domain $V$, we obtain the following expression for Lagrange functional: 
\begin{equation}\label{lagr2}
\begin{array}{r}
{\displaystyle
    \mathcal{L} = -\frac{1}{12} \sum_{i,j=1}^m \ \int\limits_{(V)} \biggl\{ K_{ij}^{(2)} \left(\nabla n_i\left( \mathbf{r} \right)\cdot \nabla n_j\left( \mathbf{r} \right) \right) } \\ 
{\displaystyle  +  Q_{ij}^{(2)} \sum_{\alpha=1}^3 
\left( \nabla \left[ D_i^{\alpha} \left( \mathbf{r} \right) n_i \left( \mathbf{r} \right) \right] \cdot \nabla \left[ D_j^{\alpha} \left( \mathbf{r} \right) n_j \left( \mathbf{r} \right) \right] \right) } \\ 
{\displaystyle  +  R_{ij}^{(2)} \sum_{\alpha=1}^3 
\left( \nabla \left[ M_i^{\alpha} \left( \mathbf{r} \right) n_i \left( \mathbf{r} \right) \right] \cdot \nabla \left[ M_j^{\alpha} \left( \mathbf{r} \right) n_j \left( \mathbf{r} \right) \right] \right)} \\
{\displaystyle  + 2 S_{ij}^{(2)} \sum_{\alpha=1}^3 
\left( \nabla \left[ D_i^{\alpha} \left( \mathbf{r} \right) n_i \left( \mathbf{r} \right) \right] \cdot \nabla \left[ M_j^{\alpha} \left( \mathbf{r} \right) n_j \left( \mathbf{r} \right) \right] \right) \biggr\} d\mathbf{r} } \\
{\displaystyle + \frac{1}{2} \sum_{i,j=1}^m\ \int\limits_{(V)} \biggl[ K_{ij}^{(0)} + Q_{ij}^{(0)} \left( \mathbf{D}_i \left( \mathbf{r} \right) \cdot \mathbf{D}_j \left( \mathbf{r} \right) \right) + R_{ij}^{(0)} \left( \mathbf{M}_i \left( \mathbf{r} \right) \cdot \mathbf{M}_j \left( \mathbf{r} \right) \right) } \\
{\displaystyle + 2 S_{ij}^{(0)} \left( \mathbf{D}_i \left( \mathbf{r} \right) \cdot \mathbf{M}_j \left( \mathbf{r} \right) \right) \biggr] n_i \left( \mathbf{r} \right)\, n_j \left( \mathbf{r} \right)\,d\mathbf{r} }
 \\
{\displaystyle + \sum_{i=1}^m\ \int\limits_{(V)} \left[\left( \mathbf{E} \left(\mathbf{r} \right)\cdot \mathbf{D}_i \left(\mathbf{r} \right)  \right) + \left( \mathbf{H} \left(\mathbf{r} \right)\cdot \mathbf{M}_i \left(\mathbf{r} \right)  \right) + T\, \ln \left( \frac{n_i\left( \mathbf{r} \right)}{n\left( \mathbf{r} \right)}  \right) \right] n_i\left( \mathbf{r} \right)\, d\mathbf{r}  }\\
{\displaystyle  - \sum_{i=1}^{m}\mu_i\left[ \int\limits_{(V)}\
    n_i(\mathbf{r})\ d\mathbf{r}-N_i \right]
- \sum_{i=1}^m\ \int\limits_{(V)} \frac{\lambda_i\left( \mathbf{r} \right)}{2} \left[ \left(\mathbf{D}_i\left(\mathbf{r}  \right)\right)^2 - D_i^2 \right]\, n_i^2 \left( \mathbf{r} \right), d\mathbf{r}\ }\\ 
{\displaystyle -
\sum_{i=1}^m\ \int\limits_{(V)} \frac{\nu_i\left( \mathbf{r} \right)}{2} \left[ \left(\mathbf{M}_i\left(\mathbf{r}  \right)\right)^2 - M_i^2 \right]\, n_i^2 \left( \mathbf{r} \right) \,d\mathbf{r}  - \int\limits_{(V)}\Psi(\mathbf{r}) \left(\sum_{i=0}^m\ \omega_i\
    n_i(\mathbf{r})-1 \right) d\mathbf{r}, }
\end{array}
\end{equation}
where 
\begin{equation}\label{Kij(p)}
    K_{ij}^{(p)} = \int\limits_{(V)}\, K_{ij} \left( \mathbf{r} \right)\,  \left| \mathbf{r} \right|^p\, d\mathbf{r}, \qquad (p=0,\, 2),
\end{equation}
are the integral characteristics of the inter-atomic potentials; the parameters $Q_{ij}^{(p)}$, $R_{ij}^{(p)}$, $S_{ij}^{(p)}$ are defined similarly.

Functional~(\ref{lagr2}) is similar to Ginzburg-Landau and Cahn-Hilliard (GLCH) functionals, but in contrast to GLCH the integrand in~(\ref{lagr2}) is not a polynomial over order parameters and all the parameters have clear physical sense, due to their explicit connections~(\ref{Kij(p)}) with inter-atomic potentials. This functional is the sum of two terms
\begin{equation}\label{L=L1+L2}
    \mathcal{L} = \mathcal{L}_1 + \mathcal{L}_2,
\end{equation}
\begin{equation}\label{L1}
     {\mathcal{L}_1} = \sum_{i=1}^m \left[ \mu_i N_i + \frac{D_i^2}{2} \int\limits_{(V)} \lambda_i(\mathbf{r})\, n_i^2\left(\mathbf{r} \right)\, d\mathbf{r} + \frac{M_i^2}{2} \int\limits_{(V)} \nu_i(\mathbf{r})\, n_i^2\left(\mathbf{r} \right)\, d\mathbf{r} \right] + \int\limits_{(V)} \Psi(\mathbf{r}) d\mathbf{r}
\end{equation}
is the part of the functional~(\ref{lagr2}), which does not depend in explicit form on functions $\mathbf{D}_i(\mathbf{r})$, $\mathbf{M}_i(\mathbf{r})$, ${n}_i(\mathbf{r})$, and
\begin{equation}\label{L-2}
\begin{array}{r}
{\displaystyle
    \mathcal{L}_2 = -\frac{1}{12} \sum_{i,j=1}^m \ \int\limits_{(V)} \biggl\{ K_{ij}^{(2)} \left(\nabla n_i\left( \mathbf{r} \right)\cdot \nabla n_j\left( \mathbf{r} \right) \right) 
} \\ 
{\displaystyle  
+  Q_{ij}^{(2)} \sum_{\alpha=1}^3 
\left( \nabla \left[ D_i^{\alpha} \left( \mathbf{r} \right) n_i \left( \mathbf{r} \right) \right] \cdot \nabla \left[ D_j^{\alpha} \left( \mathbf{r} \right) n_j \left( \mathbf{r} \right) \right] \right) } \\ 
{\displaystyle  +  R_{ij}^{(2)} \sum_{\alpha=1}^3 
\left( \nabla \left[ M_i^{\alpha} \left( \mathbf{r} \right) n_i \left( \mathbf{r} \right) \right] \cdot \nabla \left[ M_j^{\alpha} \left( \mathbf{r} \right) n_j \left( \mathbf{r} \right) \right] \right)} \\
{\displaystyle  + 2 S_{ij}^{(2)} \sum_{\alpha=1}^3 
\left( \nabla \left[ D_i^{\alpha} \left( \mathbf{r} \right) n_i \left( \mathbf{r} \right) \right] \cdot \nabla \left[ M_j^{\alpha} \left( \mathbf{r} \right) n_j \left( \mathbf{r} \right) \right] \right) \biggr\} d\mathbf{r} } \\
{\displaystyle + \frac{1}{2} \sum_{i,j=1}^m\ \int\limits_{(V)} \biggl[ K_{ij}^{(0)} + Q_{ij}^{(0)} \left( \mathbf{D}_i \left( \mathbf{r} \right) \cdot \mathbf{D}_j \left( \mathbf{r} \right) \right) + R_{ij}^{(0)} \left( \mathbf{M}_i \left( \mathbf{r} \right) \cdot \mathbf{M}_j \left( \mathbf{r} \right) \right) } \\
{\displaystyle + 2 S_{ij}^{(0)} \left( \mathbf{D}_i \left( \mathbf{r} \right) \cdot \mathbf{M}_j \left( \mathbf{r} \right) \right) \biggr] n_i \left( \mathbf{r} \right)\, n_j \left( \mathbf{r} \right)\,d\mathbf{r} }
 \\
{\displaystyle + \sum_{i=1}^m\ \int\limits_{(V)} \left[\left( \mathbf{E} \left(\mathbf{r} \right)\cdot \mathbf{D}_i \left(\mathbf{r} \right)  \right) + \left( \mathbf{H} \left(\mathbf{r} \right)\cdot \mathbf{M}_i \left(\mathbf{r} \right)  \right) + T\, \ln \left( \frac{n_i\left( \mathbf{r} \right)}{n\left( \mathbf{r} \right)}  \right) \right] n_i\left( \mathbf{r} \right)\, d\mathbf{r}  }\\
{\displaystyle  
 - \sum_{i=1}^m\int\limits_{(V)} \left\{ \mu_i\, n_i (\mathbf{r}) + \frac{\lambda_i(\mathbf{r})}{2}\, n_i^2 (\mathbf{r})\, \mathbf{D}_i^2(\mathbf{r}) +  \frac{\nu_i(\mathbf{r})}{2}\, n_i^2 (\mathbf{r})\, \mathbf{M}_i^2(\mathbf{r}) + \Psi(\mathbf{r})\,\omega_i\, n_i(\mathbf{r}) \right\}\,d\mathbf{r}
}\\
\end{array}
\end{equation}
is the second part, depending on $n_i(\mathbf{r})$,  ${D}_i^{\alpha}(\mathbf{r})$, ${M}_i^{\alpha}(\mathbf{r})$ in explicit form.

The term $\mathcal{L}_1$ does not give any contribution into derivatives of the functional~(\ref{lagr2}) with respect to $n_i(\mathbf{r})$,  ${D}_i^{\alpha}(\mathbf{r})$, ${M}_i^{\alpha}(\mathbf{r})$. Therefore, the variational problem for the functional~(\ref{lagr2}) is equivalent to the variational problem for the reduced functional $\mathcal{L}_2 $  with additional conditions~(\ref{packing}), (\ref{N}), (\ref{D-M}). This reduced functional has a form 
\begin{equation}\label{L3}
\mathcal{L}_2 \left(  \left\{u_s(\mathbf{r})  \right\} \right) =  \int\limits_{(V)} \Lambda\left( u_s(\mathbf{r}),\, \nabla u_s(\mathbf{r})  \right)\, d\mathbf{r},
\end{equation}
where $u_s(\mathbf{r})$ denotes all the functions $n_i(\mathbf{r})$,  ${D}_i^{\alpha}(\mathbf{r})$, ${M}_i^{\alpha}(\mathbf{r})$,  and
\begin{equation}\label{Lambda}
\begin{array}{r}
{\displaystyle \Lambda\left( u_s(\mathbf{r}),\, \nabla u_s(\mathbf{r})  \right) =  - \frac{1}{12} \sum_{i,j=1}^m \biggl[
K_{ij}^{(2)} \left(\nabla n_i(\mathbf{r})\cdot \nabla
n_j(\mathbf{r}) \right) 
} \\
{\displaystyle 
+  Q_{ij}^{(2)} \sum_{\alpha=1}^3 \left( \nabla \left[ {D}^{\alpha}_i(\mathbf{r}) n_i(\mathbf{r}) \right] \cdot  \nabla \left[
{D}^{\alpha}_j (\mathbf{r}) n_j(\mathbf{r}) \right] \right)  } \\
{\displaystyle + R_{ij}^{(2)} \,
\sum_{\alpha=1}^3 \left( \nabla \left[ {M}^{\alpha}_i(\mathbf{r}) n_i(\mathbf{r}) \right] \cdot  \nabla \left[
{M}^{\alpha}_j (\mathbf{r}) n_j(\mathbf{r}) \right] \right)}\\ 
{\displaystyle + 2 S_{ij}^{(2)} \,
\sum_{\alpha=1}^3 \left( \nabla \left[ {D}^{\alpha}_i(\mathbf{r}) n_i(\mathbf{r}) \right] \cdot  \nabla \left[
{M}^{\alpha}_j (\mathbf{r}) n_j(\mathbf{r}) \right] \right)\biggr] }\\ 
{\displaystyle + \frac{1}{2} \sum_{i,j=1}^m \biggl[ K_{ij}^{(0)}
 + Q_{ij}^{(0)} \left(  \mathbf{D}_i(\mathbf{r})\cdot \mathbf{D}_j(\mathbf{r})
 \right) + R_{ij}^{(0)} \left(  \mathbf{M}_i(\mathbf{r}) \cdot \mathbf{M}_j(\mathbf{r})
 \right) } \\
{\displaystyle
+ 2 S_{ij}^{(0)} \left(  \mathbf{D}_i(\mathbf{r}) \cdot \mathbf{M}_j(\mathbf{r})
 \right) \biggr] \, n_i(\mathbf{r})\,  n_j(\mathbf{r})  }\\ 
{\displaystyle + \sum_{i=1}^m \left[ \left\{  \left( \mathbf{E}(\mathbf{r}) \cdot
\mathbf{D}_i(\mathbf{r}) \right) + \left( \mathbf{H}(\mathbf{r}) \cdot
\mathbf{M}_i(\mathbf{r}) \right) \right\} n_i(\mathbf{r}) + \frac{n_i^2(\mathbf{r})}{2} \left[ {\lambda_i(\mathbf{r})} \mathbf{D}_i^2 (\mathbf{r}) + {\nu_i(\mathbf{r})} \mathbf{M}_i^2 (\mathbf{r})  \right] \right] }\\ 
{\displaystyle + \sum_{i=1}^m \left[ T\, \ln
\left(\frac{n_i(\mathbf{r})}{n(\mathbf{r})}\right) -  \mu_i - \Psi(\mathbf{r})\, \omega_i \right]  n_i (\mathbf{r}).}
\end{array}
\end{equation}

Thus, the equilibrium distributions of the components and their electric and magnetic moments obey the Lagrange-Euler system of equations for functional~(\ref{L3})
\begin{equation}\label{Lag-Eul}
    \frac{\partial \Lambda}{\partial u_s (\mathbf{r})}\ - \ \left(  \nabla\, \cdot  \frac{\partial \Lambda}{\partial \left(  \nabla u_s (\mathbf{r})  \right)} \right) \, = \, 0,
\end{equation}
together with conditions~(\ref{packing}), (\ref{N}), (\ref{D-M}).

Substitution~(\ref{Lambda}) into (\ref{Lag-Eul}) with $u_s(\mathbf{r}) = n_i(\mathbf{r})$, $u_s(\mathbf{r}) = D_i^{\alpha}(\mathbf{r})$, $u_s(\mathbf{r}) = M_i^{\alpha}(\mathbf{r})$ leads to following equations:
\begin{equation}\label{L-ni}
\begin{array}{r}
     {\displaystyle  \sum_{j=1}^m \biggl[ K_{ij}^{(0)}
 + Q_{ij}^{(0)} \left(  \mathbf{D}_i(\mathbf{r})\cdot \mathbf{D}_j(\mathbf{r})
 \right) + R_{ij}^{(0)} \left(  \mathbf{M}_i(\mathbf{r}) \cdot \mathbf{M}_j(\mathbf{r})
 \right)
+ S_{ij}^{(0)} \left(  \mathbf{D}_i(\mathbf{r}) \cdot \mathbf{M}_j(\mathbf{r})
 \right) \biggr] \,  n_j(\mathbf{r}) } \\
{\displaystyle + \sum_{j=1}^m \sum_{\alpha=1}^3 \biggl[  Q_{ij}^{(2)} \left( \nabla   {D}_i^{\alpha}(\mathbf{r})\cdot \nabla \left[ {D}_j^{\alpha}(\mathbf{r}) n_j(\mathbf{r}) \right] \right) + R_{ij}^{(2)} \left( \nabla   {M}_i^{\alpha}(\mathbf{r})\cdot \nabla \left[ {M}_j^{\alpha}(\mathbf{r}) n_j(\mathbf{r}) \right] \right)} \\
{\displaystyle
+ \left[  S_{ij}^{(2)} \left( \nabla   {D}_i^{\alpha}(\mathbf{r})\cdot \nabla \left[ {M}_j^{\alpha}(\mathbf{r}) n_j(\mathbf{r}) \right] \right) +  S_{ij}^{(2)} \left( \nabla   {M}_i^{\alpha}(\mathbf{r})\cdot \nabla \left[ {D}_j^{\alpha}(\mathbf{r}) n_j(\mathbf{r}) \right] \right)  \right] \biggr] }\\
{\displaystyle + \biggl[ \left( \mathbf{E}(\mathbf{r}) \cdot
\mathbf{D}_i(\mathbf{r}) \right)  + \left( \mathbf{H}(\mathbf{r}) \cdot
\mathbf{M}_i(\mathbf{r}) \right) + n_i \left( \mathbf{r} \right) \left[ {\lambda_i(\mathbf{r})} \mathbf{D}_i^2 (\mathbf{r}) + {\nu_i(\mathbf{r})} \mathbf{M}_i^2 (\mathbf{r}) \right]} \\
{\displaystyle
+ T\, \ln
\left(\frac{n_i(\mathbf{r})}{n(\mathbf{r})}\right) \biggr] -\mu_i + \omega_i\Psi \left( \mathbf{r} \right) }\\
{\displaystyle = - \frac{1}{6} \sum_{j=1}^m \Biggl[   K_{ij}^{(2)} \Delta n_j (\mathbf{r}) + \sum_{\alpha} Q_{ij}^{(2)} \left(\nabla \cdot \left\{ D_i^{\alpha}(\mathbf{r})\,\nabla \left(D_j^{\alpha}(\mathbf{r})\, n_j (\mathbf{r})  \right)  \right\}  \right)  }\\
{\displaystyle + \sum_{\alpha} R_{ij}^{(2)} \left(\nabla \cdot \left\{ M_i^{\alpha}(\mathbf{r})\,\nabla \left(M_j^{\alpha}(\mathbf{r})\, n_j (\mathbf{r})  \right)  \right\}  \right)  }\\
{\displaystyle +  \sum_{\alpha=1}^3 S_{ij}^{(2)} \biggl\{ \left(\nabla \cdot \left\{ D_i^{\alpha}(\mathbf{r})\,\nabla \left(M_j^{\alpha}(\mathbf{r})\, n_j (\mathbf{r})  \right)  \right\}  \right) + \left(\nabla \cdot \left\{ M_i^{\alpha}(\mathbf{r})\,\nabla \left(D_j^{\alpha}(\mathbf{r})\, n_j (\mathbf{r})  \right)  \right\}  \right)  \biggr\} \Biggr]};
\end{array}
\end{equation}
\begin{equation}\label{L-D}
\begin{array}{r}
{\displaystyle \sum_{j=1}^m \left\{ Q_{ij}^{(0)}\, D_j^{\alpha}(\mathbf{r}) +  S_{ij}^{(0)}\, M_j^{\alpha}(\mathbf{r})  \right\} n_i(\mathbf{r})\, n_j(\mathbf{r}) + E^{\alpha}(\mathbf{r})\, n_i(\mathbf{r}) + \lambda_i(\mathbf{r}) n_i^2  \left( \mathbf{r} \right) D_i^{\alpha}(\mathbf{r}) }\\
{\displaystyle -\frac{1}{6}\, \sum_{j=1}^m \left\{Q_{ij}^{(2)} \left( \nabla n_i(\mathbf{r}) \cdot \nabla \left[D_j^{\alpha}(\mathbf{r}) n_j(\mathbf{r})   \right] \right) +  S_{ij}^{(2)} \left( \nabla n_i(\mathbf{r}) \cdot \nabla \left[M_j^{\alpha}(\mathbf{r}) n_j(\mathbf{r})   \right] \right)  \right\}} \\
{\displaystyle = - \frac{1}{6}\, \nabla\cdot \sum_{j=1}^m \left\{ Q_{ij}^{(2)} n_i(\mathbf{r})\, \nabla \left[ D_j^{\alpha} (\mathbf{r}) n_j(\mathbf{r}) \right]  + S_{ij}^{(2)} n_i(\mathbf{r})\, \nabla \left[ M_j^{\alpha} (\mathbf{r}) n_j(\mathbf{r}) \right]  \right\} };
\end{array}
\end{equation}
\begin{equation}\label{L-M}
\begin{array}{r}
{\displaystyle \sum_{j=1}^m \left\{ R_{ij}^{(0)}\, M_j^{\alpha}(\mathbf{r}) + \frac{1}{2}\, S_{ij}^{(0)}\, D_j^{\alpha}(\mathbf{r})  \right\} n_i(\mathbf{r})\, n_j(\mathbf{r}) + H^{\alpha}(\mathbf{r})\, n_i(\mathbf{r}) + \nu_i(\mathbf{r}) n_i^2  \left( \mathbf{r} \right) M_i^{\alpha}(\mathbf{r}) }\\
{\displaystyle -\frac{1}{6}\, \sum_{j=1}^m \left\{R_{ij}^{(2)} \left( \nabla n_i(\mathbf{r}) \cdot \nabla \left[M_j^{\alpha}(\mathbf{r}) n_j(\mathbf{r})   \right] \right) + \frac{1}{2}\, S_{ij}^{(2)} \left( \nabla n_i(\mathbf{r}) \cdot \nabla \left[D_j^{\alpha}(\mathbf{r}) n_j(\mathbf{r})   \right] \right)  \right\} } \\
{\displaystyle = - \frac{1}{6}\, \nabla\cdot \sum_{j=1}^m \left\{ R_{ij}^{(2)} n_i(\mathbf{r})\, \nabla \left[ M_j^{\alpha} (\mathbf{r}) n_j(\mathbf{r}) \right]  + S_{ij}^{(2)} n_i(\mathbf{r})\, \nabla \left[ D_j^{\alpha} (\mathbf{r}) n_j(\mathbf{r}) \right]  \right\} },
\end{array}
\end{equation}
respectively.

The system of equations (\ref{L-ni}, \ref{L-D}, \ref{L-M}) with conditions~(\ref{packing}, \ref{N}, \ref{D-M}) describes space distributions of the components and local electric and magnetic moments for the case of {\em full thermodynamic equilibrium} in the system.

\subsection{Partial equilibrium in the system}
Complete thermodynamic equilibrium in condensed systems can be realized, as a rule, on astronomical scales of time only. Suppose that the local magnetization and the local electric polarization in solid-state structures are the fast variables while the space distributions of the components are the slow variables. Then the magnetizations and the electric polarizations can be considered on appropriate time scales as equilibrium functions while the space distributions of the components are frozen variables. There are many examples of such systems. In particular, it could be considered a layered system such as multiferroic magnetoelectric composites\cite{Bich1,Bich2}. For this situation the functions $n_i  \left( \mathbf{r} \right) $  are the fixed by the sample preparation technology while the functions $\mathbf{D}_i \left(\mathbf{r} \right) $   and $\mathbf{M}_i \left(\mathbf{r} \right) $ obey the equations~(\ref{L-D}) and~(\ref{L-M}) with conditions~(\ref{D-M}).

Let us consider as example a model of two-component system. Let the first component particles have the electric moment $D$  and second component particles have magnetic moment $M$. Space distribution of these components is prescribed by their local densities $n_1 \left( \mathbf{r} \right)$ and $n_2 \left( \mathbf{r} \right) $, respectively. The functional~(\ref{L-2}) for this case has the following form:
\begin{equation}\label{L2-DM}
\begin{array}{r}
    {\displaystyle  \mathcal{L}_2 \left(  \left\{\mathbf{D} \left( \mathbf{r} \right) \right\},\,\left\{\mathbf{M} \left( \mathbf{r} \right) \right\}  \right) = -\frac{1}{12}\sum_{\alpha=1}^3\,\int\limits_{(V)}\, \biggl\{ Q^{(2)}  \left( \nabla \left[D^{\alpha} \left(\mathbf{r} \right)n_1 \left(\mathbf{r} \right) \right] \cdot \nabla \left[D^{\alpha} \left(\mathbf{r} \right)n_1 \left(\mathbf{r} \right) \right] \right) }\\
{\displaystyle + R^{(2)} \left( \nabla \left[M^{\alpha} \left(\mathbf{r} \right)n_1 \left(\mathbf{r} \right) \right] \cdot \nabla \left[M^{\alpha} \left(\mathbf{r} \right)n_1 \left(\mathbf{r} \right) \right] \right) }\\
{\displaystyle
 + 2  S^{(2)}  \left( \nabla \left[D^{\alpha} \left(\mathbf{r} \right)n_1 \left(\mathbf{r} \right) \right] \cdot \nabla \left[M^{\alpha} \left(\mathbf{r} \right)n_1 \left(\mathbf{r} \right) \right] \right) \biggr\}d\mathbf{r}}\\
{\displaystyle + \frac{1}{2} \int\limits_{(V)} \biggl[Q^{(0)}  \left(\mathbf{D}(\mathbf{r}) \right)^2 \left(n_1(\mathbf{r}) \right)^2 + R^{(0)}  \left(\mathbf{M}(\mathbf{r}) \right)^2 \left(n_2(\mathbf{r}) \right)^2 
 }\\
{\displaystyle + 2 S^{(0)}  \left(\mathbf{D}(\mathbf{r}) \cdot \mathbf{M}(\mathbf{r}) \right) n_1(\mathbf{r}) n_2(\mathbf{r}) \biggr] d\mathbf{r} }\\
{\displaystyle
 + \int\limits_{(V)} \biggl[  \left(\mathbf{D}(\mathbf{r}) \cdot \mathbf{E}(\mathbf{r}) \right) n_1(\mathbf{r})  + \left(\mathbf{M}(\mathbf{r}) \cdot \mathbf{H}(\mathbf{r}) \right) n_2(\mathbf{r}) }\\
{\displaystyle - \frac{\lambda(\mathbf{r})}{2} n_1^2(\mathbf{r})  \left(\mathbf{D}(\mathbf{r}) \right)^2 - \frac{\nu(\mathbf{r})}{2} n_2^2(\mathbf{r})  \left(\mathbf{M}(\mathbf{r}) \right)^2 \biggr]d\mathbf{r},}
\end{array}
\end{equation}
where $Q^{(p)} = Q_{11}^{(p)}$, $R^{(p)} = R_{22}^{(p)}$, $S^{(p)} = S_{12}^{(p)}$, ($p = 0, \,2$).

Equilibrium distributions of electric and magnetic moments at given densities $n_1(\mathbf{r})$ and $n_2(\mathbf{r})$ obey the equations
\begin{equation}\label{DM}
    \left\{
\begin{array}{l}
    {\displaystyle Q^{(0)} n_1^2(\mathbf{r}) D^{\alpha}(\mathbf{r}) + S^{(0)} n_1(\mathbf{r}) n_2(\mathbf{r}) M^{\alpha}(\mathbf{r}) + E^{\alpha}(\mathbf{r}) n_1(\mathbf{r}) - \lambda(\mathbf{r}) n_1^2(\mathbf{r}) D^{\alpha}(\mathbf{r})}\\
{\displaystyle - \frac{1}{6} \left\{Q^{(2)}  \left( \nabla n_1(\mathbf{r}) \cdot \nabla \left[ D^{\alpha}(\mathbf{r}) n_1(\mathbf{r})\right] \right) + S^{(2)}  \left( \nabla n_1(\mathbf{r}) \cdot \nabla \left[ M^{\alpha}(\mathbf{r}) n_2(\mathbf{r})\right] \right) \right\}  }\\
{\displaystyle = - \frac{1}{6} \nabla \cdot \left\{Q^{(2)} n_1(\mathbf{r})  \nabla \left[ D^{\alpha}(\mathbf{r}) n_1(\mathbf{r})\right]  + S^{(2)} n_1(\mathbf{r}) \nabla \left[ M^{\alpha}(\mathbf{r}) n_2(\mathbf{r})\right] \right\}};\\ \\
    {\displaystyle R^{(0)} n_2^2(\mathbf{r}) M^{\alpha}(\mathbf{r}) + S^{(0)} n_1(\mathbf{r}) n_2(\mathbf{r}) D^{\alpha}(\mathbf{r}) + H^{\alpha}(\mathbf{r}) n_2(\mathbf{r}) - \nu(\mathbf{r}) n_2^2(\mathbf{r}) M^{\alpha}(\mathbf{r})}\\
{\displaystyle - \frac{1}{6} \left\{R^{(2)}  \left( \nabla n_2(\mathbf{r}) \cdot \nabla \left[ M^{\alpha}(\mathbf{r}) n_2(\mathbf{r})\right] \right) + S^{(2)}  \left( \nabla n_2(\mathbf{r}) \cdot \nabla \left[ D^{\alpha}(\mathbf{r}) n_1(\mathbf{r})\right] \right) \right\}  }\\
{\displaystyle = - \frac{1}{6} \nabla \cdot \left\{R^{(2)} n_2(\mathbf{r})  \nabla \left[ M^{\alpha}(\mathbf{r}) n_2(\mathbf{r})\right]  + S^{(2)} n_2(\mathbf{r}) \nabla \left[ D^{\alpha}(\mathbf{r}) n_1(\mathbf{r})\right] \right\}}.
\end{array}
\right.
\end{equation}
These equations can be slightly simplified. With account of the conditions~(\ref{D-M}) the complete system for equilibrium distribution of the electric and magnetic moments has the following form:
\begin{equation}\label{DM2}
\left\{
\begin{array}{l}
{\displaystyle \left[-\frac{1}{6} Q^{(2)} \Delta - Q^{(0)} + \lambda(\mathbf{r}) \right] u^{\alpha}(\mathbf{r}) +  \left[- \frac{1}{6} S^{(2)} \Delta - S^{(0)} \right] v^{\alpha}(\mathbf{r}) - E^{\alpha}(\mathbf{r}) = 0;}\\
{\displaystyle \left[-\frac{1}{6} S^{(2)} \Delta - S^{(0)}\right] u^{\alpha}(\mathbf{r}) +  \left[- \frac{1}{6} R^{(2)} \Delta - R^{(0)} + \nu(\mathbf{r}) \right] v^{\alpha}(\mathbf{r}) - H^{\alpha}(\mathbf{r}) = 0;}\\
{\displaystyle \sum_{\alpha=1}^3  \left(u^{\alpha}(\mathbf{r}) \right)^2 - D^2 n_1^2 (\mathbf{r}) =0; }\\
{\displaystyle \sum_{\alpha=1}^3  \left(v^{\alpha}(\mathbf{r}) \right)^2 - M^2 n_2^2 (\mathbf{r}) =0,}    
\end{array}
\right.
\end{equation}
where $\Delta$ is the Laplace operator, and
\begin{equation}\label{uv}
\left\{
\begin{array}{l}
    {\displaystyle u^{\alpha}(\mathbf{r}) = D^{\alpha}(\mathbf{r})n_1(\mathbf{r})}\\
    {\displaystyle v^{\alpha}(\mathbf{r}) = M^{\alpha}(\mathbf{r})n_2(\mathbf{r}).}
\end{array}
\right.
\end{equation}

The first and second equations in~(\ref{DM2}) can be written in the matrix form:
\begin{equation}\label{Schr1}
 \left(
\begin{array}{@{}cc@{}}
\left[-\frac{1}{6} Q^{(2)} \Delta - Q^{(0)} + \lambda(\mathbf{r}) \right] & \left[- \frac{1}{6} S^{(2)} \Delta - S^{(0)} \right]\\ \\
\left[-\frac{1}{6} S^{(2)} \Delta - S^{(0)}\right] & \left[- \frac{1}{6} R^{(2)} \Delta - R^{(0)} + \nu(\mathbf{r}) \right]
\end{array}
 \right)
\begin{pmatrix} u^{\alpha}(\mathbf{r})\\ \\ u^{\alpha}(\mathbf{r})
\end{pmatrix} = 
\begin{pmatrix} E^{\alpha}(\mathbf{r})\\ \\ H^{\alpha}(\mathbf{r})
\end{pmatrix}. 
\end{equation}
In absence of the external fields this matrix equation is fully similar to the coupled Schr\"dinger equations for functions $u^{\alpha}(\mathbf{r})$ and $u^{\alpha}(\mathbf{r})$. The last two equations in~(\ref{DM2}) are similar to the wave functions normalization; they make possible to find the unknown functions $\lambda(\mathbf{r})$ and $\nu(\mathbf{r})$  in just the same way as the normalization condition in quantum mechanics permits to find the spectrum. Thus, classical statistical thermodynamics of the quasi-equilibrium condensed systems with internal degrees of freedom can be based on the standard quantum mechanical methods.

\section{Conclusion}
The paper contains a development of the generalized lattice model in following ways.
\begin{enumerate}
    \item The separation of the short-range inter-atomic repulsions and the smooth long-range  parts of the inter-atomic potentials. Short-range repulsions take into account by means of packing condition, long-range parts take into account in effective field approximation.
    \item Dependence of the inter-atomic potentials on the atomic internal degrees of freedoms (such as atomic electric and/or magnetic momentum) taken into account.
    \item The Helmholtz free energy functional in the generalized lattice model is reduced to the Ginzburg-Landau-Cahn-Hilliard-like form. The connection between the inter-atomic potentials characteristics and the parameters of the GLCH-like functional is obtained.
    \item The equations for both equilibrium and quasi-equilibrium states in condensed systems are derived. It is shown that equilibrium distribution of the fast internal degrees of freedom by frozen space distribution of the components obeys to the Schr\"odinger-like equation.
\end{enumerate}

One of the most interesting applications of the GLM is research of layered structures with alternation of magnetic and ferroelectric layers~\cite{Bich1,Bich2}. In particular, this approach permits
\begin{itemize}
    \item to take into account lattices misfits on the interphases boundaries without any auxiliary assumptions;
    \item to impart the physical interpretation to the phenomenological models like GLCH models;
    \item to find the ways for prognosis of the layered structures.
\end{itemize}

\section*{Acknowledgments}

The work was partially supported by the Program of Russian Ministry of Education and Science.

%------------------------------------------------------------------------------
\end{document}